# CCD Photometry of Nova V1500 Cygni Twenty Years After

by

I. S e m e n i u k,  A. O l e c h  and  M. N a l e ż y t y

Warsaw University Observatory, Al. Ujazdowskie 4, 00-478 Warszawa, Poland
e-mail:(is,olech,nalezyty)@sirius.astrouw.edu.pl

ABSTRACT

We report on CCD photometry of Nova V1500 Cygni obtained in July 1995 to show that twenty years after outburst, being of about 18 mag, the star can still be observed with small telescopes. The 0.1396 day period continues to be stable.

**Key words:** *Stars: individual: V1500 Cyg – binaries: close – novae, cataclysmic variables*

## 1. Introduction

V1500 Cygni (Nova Cygni 1975), the brightest, fastest and the most intriguing nova of the last twenty years, faded now to 18 mag. However, despite of its faintness, it still appears to be an object suitable for observations with small telescopes, and this is possible due to the CCD technique. The nova is still worthwhile to be observed. Twelve years after its outburst, polarimetric observations (Stockman, Schmidt and Lamb 1988) revealed that the object was an intermediate polar with the spin period $P_{\rm spin}$ of its white dwarf component being about 2% shorter than the orbital period $P_{\rm orb}$ equal to 0.1396 days. Subsequently, the spin period appeared to be increasing (Schmidt and Stockman 1991, Schmidt, Liebert and Stockman 1995) with the rate of increase indicating synchronization of the rotational and orbital motions in about 170 years. On this time-scale the star will be returning to its pre-outburst AM Herculis state. It is now commonly believed that – as suggested by Semeniuk *et al.* in 1977 – prior to its outburst V1500 Cyg was an AM Her type star. The 1975 explosion has broken down the spin/orbit synchronism and the quickly decreasing photometric period observed for the first two seasons after outburst was just $P_{\rm spin}$. Since 1977 the 0.1396 day orbital modulation has been dominating in optical photometry (Patterson 1979) with its full amplitude reaching now 1 mag. The rotational modulation has become so insignificant that a direct determination of its extrema from light curves is impossible. Nevertheless, the evolution of the rotational velocity of the white dwarf in the system can be traced photometrically as a change of a beat period between $P_{\rm spin}$ and $P_{\rm orb}$.

Our last observations of Nova V1500 Cygni made at the Ostrowik station with a photoelectric photometer were performed in October 1976 (Semeniuk *et al.* 1977), when the star was of about 12 mag. Afterwards the star became inaccessible for the Ostrowik photoelectric photometry.

In this paper we present new Ostrowik observations of V1500 Cygni obtained there for the first time after an about 19 year interval. Although the star faded now to 18 mag, it could be successfully observed with our 0.6-m telescope due to the CCD technique. With our observations we would like to call attention of other CCD observers on small telescopes to this faint but still interesting object.

## 2. Observations

We have observed V1500 Cygni on three nights of July 27, 28 and, 29, 1995 at the Ostrowik station of the Warsaw University Observatory with the 0.6-m telescope equipped with a Tektronics TK512 CCD camera, described by Udalski and Pych (1992).

The observations were made in white light with the exposure time of 300 s and a dead time of 20 s between frames. The data reductions have been performed with a standard procedure used at the Warsaw University Observatory based on the IRAF package. The profile photometry has been done with the DAOphotII package.

Fig. 1 presents the light curves of V1500 Cygni. $\Delta m$ denotes the difference of the instrumental magnitudes between the variable and comparison stars. The comparison was the star C1 from the finding chart of Kałużny and Semeniuk (1987). The *UBVRI* photometry performed by these authors gives for the star C1 the following parameters: $V = 15.67$, $U - B = 0.55$, $B - V = 1.00$, $V - R = 0.58$, and $R - I = 0.59$. A rough estimate of a mean $V$ magnitude of V1500 Cygni based on extrapolation of the mean $V$ magnitudes observed for the last eight years gives $V \approx 17.8$ for our epoch. From the photometry performed at Ostrowik on the night of November 26, 1995 in the filter $R$ and using the $V$ and $R$ magnitudes of Nova and the star C1 as published by Kałużny and Semeniuk (1987), we got 18.1 mag for the mean $V$ magnitude of V1500 Cygni, what agrees quite well with the value obtained from the extrapolation, if one takes into account that the mean level of Nova brightness varies from night to night by a few tenths of magnitude. The mean observational error of our present observations is equal to 0.02 mag. The shape of the light curves appears to be nearly sinusoidal with the full amplitude of about 1 mag. Contrary to the $B$ observations from 1986 (Kałużny and Semeniuk 1987) no scatter of points in extrema, indicating a flickering, is observed. The lack of flickering has already been reported earlier by DeYoung (1993) for his $V$ observations. Fig. 1 shows that the depth of minima is changing by about 0.2 mag from cycle to cycle, and also the level of maxima varies by about 0.2 mag from night to night. We presume that these changes result from an affecting influence of variability related to the beat period, which also seems to distort slightly the sinusoidal on the average shape of the light curves.

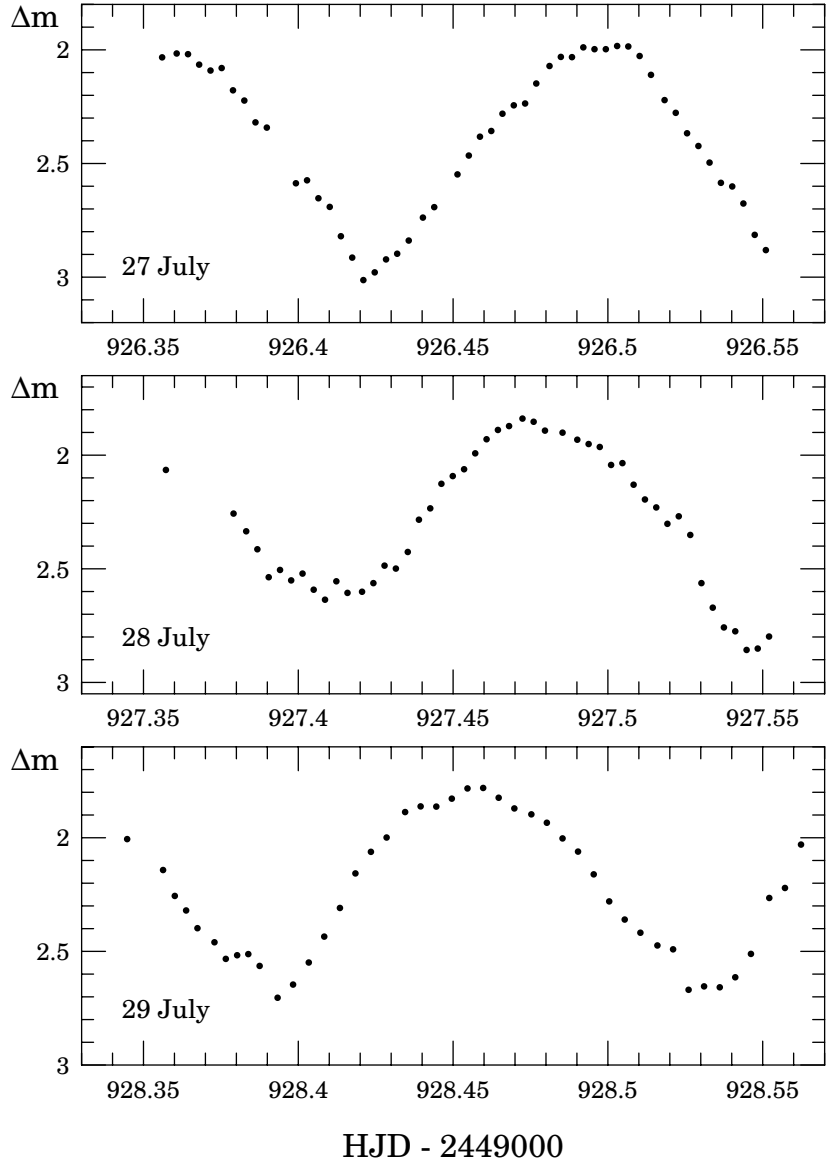

Fig. 1. White light curves of V1500 Cygni observed during three consecutive nights in July 1995. $\Delta m$ is the magnitude difference between the variable and comparison stars. The $V$ magnitude of the comparison is 15.67.

### 3. Times of Extrema and Orbital Period

>From our observations we have determined four times of minima and three times of maxima of the orbital periodicity. They are listed as the last entries in Table 1. In this table we have also collected all available times of extrema determined by earlier CCD observers of the object. Some extrema of Kałużny and Chlebowski (1988) have been redetermined using their light curves.

T a b l e 1

Times of CCD Extrema of the Orbital Periodicity of V1500 Cygni

| Minima | | | | Maxima | | | |
|---|---|---|---|---|---|---|---|
| HJD 2440000+ | E | $O - C$ | Ref. | HJD 2440000+ | E | $O - C$ | Ref. |
| 6694.6744 | 0 | 0.0014 | 1 | 6694.7424 | 0 | 0.0016 | 1 |
| 6696.6316 | 14 | 0.0040 | 1 | 6696.6830 | 14 | –0.0124 | 1 |
| 6696.7654 | 15 | –0.0018 | 1 | 6960.8370 | 1906 | –0.0061 | 2 |
| 6960.7728 | 1906 | –0.0026 | 2 | 6961.8272 | 1913 | 0.0068 | 2 |
| 6960.8996 | 1907 | –0.0154 | 2 | 6962.8025 | 1920 | 0.0049 | 2 |
| 6961.7580 | 1913 | 0.0054 | 2 | 6963.7785 | 1927 | 0.0036 | 2 |
| 6961.8952 | 1914 | 0.0029 | 2 | 7743.9310 | 7515 | –0.0010 | 3 |
| 6962.8788 | 1921 | 0.0092 | 2 | 7744.9120 | 7522 | 0.0027 | 3 |
| 6963.8359 | 1928 | –0.0109 | 2 | 9244.7770 | 18265 | 0.0059 | 4 |
| 7055.7150 | 2586 | 0.0028 | 3 | 9926.4980 | 23148 | –0.0031 | 5 |
| 7743.8710 | 7515 | 0.0065 | 3 | 9927.4752 | 23155 | –0.0032 | 5 |
| 9244.8434 | 18266 | 0.0000 | 4 | 9928.4559 | 23162 | 0.0003 | 5 |
| 9926.4232 | 23148 | –0.0107 | 5 | | | | |
| 9927.4108 | 23155 | –0.0004 | 5 | | | | |
| 9928.3938 | 23162 | 0.0053 | 5 | | | | |
| 9928.5324 | 23163 | 0.0043 | 5 | | | | |

References: 1. Kałużny and Semeniuk 1987, 2. Kałużny and Chlebowski 1988, 3. Schmidt, DeYoung and Wagner 1989, 4. DeYoung 1993, 5. This work.

The least-squares method employed independently for minima and maxima of Table 1 showed that their times are satisfactorily well described with the following linear ephemerides:

$$\text{HJD}_{\text{Min}} = 2446694.6730 + 0.13961296\,E$$
$$\pm\ 0.0024 \pm 0.00000019$$

$$\text{HJD}_{\text{Max}} = 2446694.7408 + 0.13961294\,E$$
$$\pm\ 0.0024 \pm 0.00000018$$

Columns 3 and 7 of Table 1 contain the $O - C$ residuals from the ephemerides. The orbital period has the same value as obtained by Kałużny and Semeniuk (1987) nine years ago. It is also the same, within the error limits, as calculated for the 1978–82 extrema of V1500 Cygni (Kruszewski, Semeniuk and Duerbeck 1983). It means that during the last 18 years the 0.1396 day period has been stable.

### 4. Conclusions

We have presented white light curves of Nova V1500 Cygni obtained in July 1995. They show that being now – in its minimum orbital brightness – an object of about 18.5 mag in $V$, the star can be sufficiently precisely photometered with a small telescope provided that it is equipped with a CCD camera. The amplitude of the orbital modulation is equal to 1 mag, *i.e.*, remains approximately the same since the beginning of the CCD observations of Nova in 1986.

We suggest that the changing level of maxima and minima of the light curves results from the beat phenomenon of the rotational and orbital motions. The presence of the 7.69 day period in a 1981–84 photometry of the nova has been already reported by Pavlenko and Pelt (1991). The ephemeris of Schmidt, Liebert and Stockman (1995) predicts for our epoch $P_{\rm spin} = 0.137216$ days, what gives for the beat period 7.99 days. Observations of a photometric modulation with this period and determination of phases of its extrema, could contribute to understanding of geometry of the system, and be an additional confirmation of evolution of the white dwarf rotational velocity. Due to this evolution the beat period will be progressively lengthening with time becoming more difficult for observation. It is fortunate, however, that the nova has not faded to its pre-outburst brightness – fainter than 21 mag – but stays approximately at the level of 18 mag, fading no more than a few hundredths of magnitude per year. This circumstance should be a stimulus for observers on small telescopes to include V1500 Cygni into their observational programs.

Times of extrema of the 0.1396 day modulation from the last nine years confirm the stability of the orbital period.

**Acknowledgements.** We are grateful to Professor Andrzej Kruszewski for comments on the manuscript. Thanks are also due to Mr P. Grzywacz for assistance in observations. This work was partly supported by KBN grant BST 501/A/95.

### REFERENCES

DeYoung, J.A. 1993, *IBVS*, No. 3963.
Kałużny, J., and Chlebowski, T. 1988, *Astrophys. J.*, **332**, 287.
Kałużny, J., and Semeniuk, I. 1987, *Acta Astron.*, **37**, 349.
Kruszewski, A., Semeniuk, I., and Duerbeck, H.W. 1983, *Acta Astron.*, **33**, 339.
Patterson, J. 1979, *Astrophys. J.*, **231**, 789.

Pavlenko, E.P., and Pelt, J. 1991, *Astrofizika*, **34**, 169.
Semeniuk, I., Kruszewski, A., Schwarzenberg-Czerny, A., Chlebowski, T., Mikołajewski, M., and Wołczyk, J. 1977, *Acta Astron.*, **27**, 301.
Schmidt, G.D., Liebert, J. and Stockman, H.S 1995, *Astrophys. J.*, **441**, 414.
Schmidt, G.D. and Stockman, H.S. 1991, *Astrophys. J.*, **371**, 749.
Schmidt, R.E., DeYoung, J.A. and Wagner, B.C. 1989, *IBVS*, No. 3402.
Stockman, H.S., Schmidt, G.D., and Lamb, D.Q. 1988, *Astrophys. J.*, **333**, 282.
Udalski, A. and Pych, W. 1992, *Acta Astron.*, **42**, 285.